\documentclass[a4paper,twocolumn,10pt]{article}
\usepackage[margin=0.7in]{geometry}
\usepackage{titling}
\usepackage{blindtext}

\usepackage{todonotes}
\usepackage{fancyhdr}

\usepackage{microtype}

\usepackage{hyphenat}
\usepackage{booktabs}  
\usepackage{comment}
\usepackage{enumitem}

\usepackage{navigator}
\usepackage{url}
\usepackage{graphicx}
\usepackage[export]{adjustbox}
\usepackage{float}
\usepackage{multirow}
\usepackage{longtable}
\usepackage{array}
\newcolumntype{L}{>{\raggedright\arraybackslash}p{3cm}}
\usepackage{pdfpages}
\usepackage{footnote}
\makesavenoteenv{tabular}

\begin{document}
	
	\title{\centering{Cyber security and the Leviathan}}
	
			\author{}
	{
		\author{
						{\rm Joseph Da Silva}\\
			Royal Holloway, University of London\\
			joseph.dasilva.2018@live.rhul.ac.uk
					}
		
	}

\maketitle

\begin{abstract}
Dedicated cyber-security functions are common in commercial businesses, who are confronted by evolving and pervasive threats of data breaches and other perilous security events. Such businesses are enmeshed with the wider societies in which they operate. Using data gathered from in-depth, semi-structured interviews with 15 Chief Information Security Officers, as well as six senior organisational leaders, we show that the work of political philosopher Thomas Hobbes, particularly Leviathan, offers a useful lens through which to understand the context of these functions and of cyber security in Western society. Our findings indicate that cyber security within these businesses demonstrates a number of Hobbesian features that are further implicated in, and provide significant benefits to, the wider Leviathan-esque state. These include the normalisation of intrusive controls, such as surveillance, and the stimulation of consumption. We conclude by suggesting implications for cyber-security practitioners, in particular, the reflexivity that these perspectives offer, as well as for businesses and other researchers..
\end{abstract}

\section{Introduction}\label{sec:introduction}
Cyber-security practice is increasingly recognised as more than a technological exercise. The application of sociological and political viewpoints to such practice, particularly in organisations, is becoming more and more common, e.g.,~\cite{burdonSignificanceSecuringCritical2019,stevensCyberSecurityPolitics2016}. In this paper, we build on these foundations by applying a number of lenses based on the work of Thomas Hobbes to a study of 15 Chief Information Security Officers (CISOs) and six senior organisational stakeholders representing 18 UK-based, but predominantly multinational, businesses. This work contributes to and extends cyber security scholarship by considering cyber security within business as a component of wider societal power structures. First, this research indicates that cyber security functions within businesses serve the interests of the state Leviathan. This positions those functions as indirect and possibly unwitting agents of the state, and cyber security itself as beneficial to the state and associated hegemonies. Second, it shows that cyber security functions within businesses operate as a Hobbesian form of control within the micro-societies of businesses, who are themselves mini-Leviathans. Third, it provides a novel sociological lens with which to explore cyber security within businesses and wider societies. We consider the key contribution of this research as being to provide a novel viewpoint on cyber-security practice that enables greater reflexivity and reflection for practitioners, as well as offering a pathway for future research.

Our research question, part of a wider study, was ‘what is the purpose of a CISO in a commercial organisation?’, and, through our analysis, we applied a range of different sociological concepts in order to derive meaning from our data. One of those lenses, motivated by multiple resonances within the data, was that of Hobbes. Hobbes' \textit{Leviathan}, in particular, has had a significant influence on Western political philosophy~\cite{arendtOriginsTotalitarianism2017,stevensCyberSecurityPolitics2016} and it is from this text that we develop our analytical lenses, in a similar vein to that followed by Burdon and Coles-Kemp~\cite{burdonSignificanceSecuringCritical2019} who applied Smith~\cite{smithCerberusLairBringing2005} as a lens to their study of cyber-security practitioners. Hobbes' thesis, expounded in \textit{Leviathan} and other works that follow a consistent thread, e.g.,~\cite{hobbesEnglishWorksThomas1839,hobbesCive2009}, is one of structured power and the establishment of an effective (bourgeois) society~\cite{macphersonIntroduction1985}. His philosophy is both political and moral, and influenced a number of other major philosophical works, e.g.,~\cite{rousseauSocialContract1968,lockeEssayConcerningHuman1997}. As others have pointed out, e.g.,~\cite{claassenHobbesMeetsModern2020}, there is a need for caution when applying historical concepts to modern situations without acknowledging the circumstances in which they were authored. However, given the influence Hobbesian thinking has had on modern society~\cite{arendtOriginsTotalitarianism2017}, it would be “a missed opportunity”~\cite[p. 103]{claassenHobbesMeetsModern2020} to ignore the value that can be offered by this analytical lens.\footnote{It should also be noted that extensive conversation regarding Hobbes continues in International Relations and Sociology and it is not within the scope of this paper to explore these debates. Hobbes offers a starting point into wider viewpoints from these disciplines and provides one perspective, rather than an authoritative view on modern societies.}

We are not the first to look at modern businesses through a Hobbesian lens, e.g.,~\cite{claassenHobbesMeetsModern2020, chandlerLeviathansMultinationalCorporations2005} and others have invoked Hobbes in reference to cyber security, e.g.,~\cite{hughesTreatyCyberspace2010,stevensCyberSecurityPolitics2016,kaminskiEscapingCyberState2010}. However, we believe we are the first to apply a business-as-Leviathan lens to concepts of cyber security in business and how these relate to the wider state Leviathan. By applying sociological lenses to cyber security within business, we aim to achieve, and encourage, greater reflexivity~\cite[p. 119]{cormackSociologyMassCulture2004} within both academia and practice as to both the intended and unintended consequences of such functions. 

Cyber security is inherently multi-disciplinary~\cite[p. 107]{hallCriticalVisualizationCase2015} and its sociological aspects have been explored by many scholars e.g.,~\cite{stevensCyberSecurityPolitics2016, coles-kempWhyShouldCybersecurity2018,deibertRiskingSecurityPolicies2010,shiresEnactingExpertiseRitual2018}, with direct calls being made for sociologists to research cyber security within organisations~\cite{dawsonFutureCybersecurityWorkforce2018}. Examples of sociological viewpoints being applied to cyber security within organisations include the exploration of social practices relating to cyber security within an organisation~\cite{ashendenSecurityDialoguesBuilding2016} and of trust building~\cite{flechaisDivideConquerRole2005}. Others have explored the role of the CISO, e.g.,~\cite{ashendenCISOsOrganisationalCulture2013,lanzChiefInformationSecurity2017,raiCisoCareerSurely2019}, including the importance of the social aspects of this role~\cite{hooperEmergingRoleCISO2016}, and, from a methodological perspective, argued for computer security research to be grounded in an interpretive socio-organisational paradigm~\cite{dhillonCurrentDirectionsSecurity2001} .

We begin by providing a brief conceptual grounding on Hobbes in Section~\ref{sec:grounding}. Next, we describe our methodology in Section~\ref{sec:methods} before presenting our research findings in Section~\ref{sec:findings}. We unpack these in Section~\ref{sec:discussion}, employing a number of concepts from Hobbes for the purpose of analysis and interpretation, before concluding in Section~\ref{sec:conclusion}, suggesting implications for practitioners and businesses, as well as future research directions.

\section{Conceptual grounding}\label{sec:grounding}
In this section, we briefly summarise a number of key concepts from Hobbes which are used as analytical lenses in Section \ref{sec:discussion} in order to interpret, and gain a deeper understanding of the findings in Section \ref{sec:findings}. We began with a ground-up analysis of our data, from which we identified a number of references to state power, fear and market dynamics in relation to cyber security. Subsequently, we brought these references into conversation with Hobbesian notions, due to the manifest linkages to his work, using these as a framework on which we built a deeper interpretation and derivation of meaning from the findings. Such an approach is well established in qualitative research that follows an interpretive paradigm, where new knowledge is developed through the application of existing theory to data, e.g., ~\cite{burdonSignificanceSecuringCritical2019}.

Hobbes has been influential in the very definition of security~\cite[p. 69]{kangasLeadingChangeComplex2019}. Other scholars have previously employed his work in exploring the links between cyber security and state security, e.g.,~\cite{coles-kempWhyShouldCybersecurity2018,kaminskiEscapingCyberState2010}, including aspects of surveillance, e.g.,~\cite{coles-kempWatchingYouWatching2014,baumanSnowdenRethinkingImpact2014} and cyber warfare, e.g.,~\cite{brennerCiviliansCyberwarfareConscripts2010}. Others have highlighted the importance of corporations in achieving national security, e.g.,~\cite{carrPublicprivatePartnershipsNational2016}, with some problematising this relationship, e.g.~\cite{eichensehrPublicPrivateCybersecurity2017}. The threat that cyber security may pose to national and international security has also been extensively explored by others, e.g.,~\cite{warfRelationalGeographiesCyberterrorism2016}, including the societal risks posed, e.g.,~\cite{siroliConsiderationsCyberDomain2018}, the impacts that responses to this may have on freedoms, e.g.,~\cite{nissenbaumWhereComputerSecurity2005} and the threats to state power associated with cyber security risk~\cite{carrPublicprivatePartnershipsNational2016}. The application of sociological perspectives to businesses is also well established, e.g., ~\cite{woodwardIndustrialOrganizationTheory1965,silvermanTheoryOrganisationsSociological1970,burrellSociologicalParadigmsOrganizational1987}. More recently, Geppert and Dörrenbächer~\cite{geppertPoliticsPowerMultinational2014} summarised how power relations in multinational businesses have been studied from a sociological perspective and highlighted the links to wider societal power structures. In a broader context, the application of social perspectives to risk is well established, e.g.,~\cite{beckRiskSocietyNew1992} with the impact, and use of, fear within society, e.g.,~\cite{giddensConsequencesModernity1990,beckWorldRisk2009,furediCultureFearRevisited2006} and how this supports wider power structures, e.g.,~\cite{neocleousCritiqueSecurity2008}, being a common thread.

We now briefly summarise a number of key concepts from Hobbes which form the basis of the analysis in Section~\ref{sec:discussion}.

\subsection{The state of nature}
\textit{Leviathan} was written during the English Civil War,\footnote{Although it was a development of earlier ideas~\cite{HobbesLeviathan}.} and this turmoil was a key concern of Hobbes, who believed that his political science could avoid any recurrence and achieve a lasting peace. It is premised on an argument that, without effective governance, humankind would exist in a state of war, “of every man, against every man”~\cite[p. 185]{hobbesLeviathan1985}.\footnote{\textit{Leviathan} is “covertly gendered”~\cite[p. 118]{carverMenMasculinitiesInternational2014}, with “the important actors in life [being] men\dots or very rarely\dots masculinized women”~\cite[p. 118]{carverMenMasculinitiesInternational2014}. The primacy Hobbes provides to men is clear throughout the text, and he was “writing for a male audience\dots from a male point of view”~\cite[p. 635n10]{distefanoMasculinityIdeologyPolitical1983}.} In this state, regardless of the existence of “actuall fighting”~\cite[p. 186]{hobbesLeviathan1985}, there is “continuall feare, and danger of violent death; And the life of man, solitary, poore, nasty, brutish, and short”~\cite[p. 186]{hobbesLeviathan1985}. The avoidance of this ‘state of nature’, as it is referred to, e.g.,~\cite{merriamHobbesDoctrineState1906}, is a primary motivation in the establishment of the Leviathan, which provides security against it in exchange for obedience. This is one of the key tenets of \textit{Leviathan}; citizens enter into a contract with the state in which this exchange takes place~\cite{baumgoldTrustHobbesPolitical2013}. The observance of this contract, according to Hobbes, was fundamental to achieving peace within a society, and was based on both “the absolute right of sovereigns to command\dots and the absolute duty of the people to obey”~\cite[p. 613]{bejanTeachingLeviathanThomas2010}. This obedience is consensual~\cite[p. 80]{chapmanLeviathanWritSmall1975} although citizens may suffer diminution as a result~\cite{wolinHobbesEpicTradition1970}. This contract, and the extension of Hobbes' ideas, may equate to tyranny, even totalitarianism~\cite{barkanRobertoEspositoPolitical2012,arendtOriginsTotalitarianism2017}. The Leviathan's “ultimate end is accumulation of power”~\cite[p. 180]{arendtOriginsTotalitarianism2017}. The tyrannical aspect is something that Hobbes himself does not deny and is, in fact, “proud to admit”~\cite[p. 188]{arendtOriginsTotalitarianism2017}; he is dismissive towards accusations of tyranny, labelling these as simply the protestatory responses of malcontents~\cite[p. 240]{hobbesLeviathan1985}.

While the Leviathan exists to avoid the state of nature, it benefits from the continued presence of this threat in the minds of its citizens. Without this threat, the Leviathan's power and dominion over its citizens is diminished; in order to exchange obedience for protection, there needs to be some peril, otherwise the equation is imbalanced. Security of citizens from threat is “[t]he \textit{raison d'être} of the state”~\cite[p. 181]{arendtOriginsTotalitarianism2017} (italics in original). Sovereignty is both achieved and maintained through fear~\cite{lloydHobbesMoralPolitical2020}. Hobbes describes how “that which enclineth men least to break the Laws, is Fear”~\cite[p. 343]{hobbesLeviathan1985} and, further, that fear may in fact be “the onely thing\dots that makes men keep them [i.e., laws]”~\cite[p. 343]{hobbesLeviathan1985}. He believed that “feare of some coercive Power”~\cite[p. 196]{hobbesLeviathan1985}, owned by the sovereign, was necessary in order to make citizens keep their promises~\cite{peacockObligationAdvantageHobbes2010}. Barkan, discussing Esposito~\cite{espositoBiosBiopoliticsPhilosophy2008}, describes “the sovereign’s power to expose life to death as opposing but also interlinked sides of a persistent immunitary [\textit{sic}] dynamic”~\cite[p. 89]{barkanRobertoEspositoPolitical2012}. In other words, it is beneficial for the sovereign for threat to life to exist, so that the sovereign can offer protection to the citizenry from such a threat; if this threat ceases to exist, or ceases to be \textit{perceived} to exist, then the power of the sovereign in commanding obedience is diminished.

\paragraph{Permanent emergency, warfare and power.}
Within the domain of International Relations (IR), the concept of permanent emergency has been established and discussed by a number of scholars, e.g.,~\cite{neocleousCritiqueSecurity2008}. This refers to the perpetuation of a state of threat, whereby a population's security, and often its way of life, are subject to, or positioned as being subject to, various forms of continuing menace. This environment facilitates the establishment of various responses to those threats that restrict the freedoms of citizens, in the name of ‘security’~\cite{neocleousCritiqueSecurity2008,bubandtVernacularSecurityPolitics2005}, a concept which has many parallels with Hobbes' state of nature.

In a Hobbesian society, there is a never-ending need for the state to expand its power; “only by constantly extending its authority and only through the process of power accumulation can it remain stable”~\cite[p. 184]{arendtOriginsTotalitarianism2017}. If such a society were to achieve “complete security”, then the state's power would crumble~\cite[p. 184]{arendtOriginsTotalitarianism2017}. Therefore, there is a need for the continual provision of “new props from the outside”~\cite[p. 184]{arendtOriginsTotalitarianism2017}, such as novel threats. The “ever-present possibility of war guarantees the Commonwealth a prospect of permanence because it makes it possible for the state to increase its power at the expense of other states”~\cite[p. 184-5]{arendtOriginsTotalitarianism2017}. According to Hobbes, “[i]t is rationally required to seek peace, but when peace is unattainable it is rationally allowed to wage war”~\cite[p. 245]{gertHobbesReason2001}. Arendt further elaborates the need for a “never-ending accumulation of power [as being] necessary for the protection of a never-ending accumulation of capital”~\cite[p. 186]{arendtOriginsTotalitarianism2017} and how this has underpinned imperialism and indeed modern society.\footnote{Modern globalisation practices being equivalent with imperialistic ones~\cite{chilcoteGlobalizationImperialism2002}. Hobbes himself was actively involved in colonial enterprise~\cite{jessenStateCompanyCorporations2012}.}

Although a Hobbesian state may hold supreme power, its citizens still hold the right to rebel against it if such rebellion is for self-defence~\cite{williamsHobbesInternationalRelations1996}. As a precaution against such an eventuality, the Hobbesian state must “have recourse to arms to enforce civil order”~\cite[p. 221]{williamsHobbesInternationalRelations1996}. In \textit{De Cive}, Hobbes describes how “[a]ll judgement therefore in a City belongs to him who hath the swords”~\cite[p. 48]{hobbesCive2009}. But the state must also remain trusted by its citizens, particularly in its determination of what is and is not a threat~\cite{williamsHobbesInternationalRelations1996}, and the most important control that the sovereign should have is over “language (which defines what is)”~\cite[p. 219-220]{williamsHobbesInternationalRelations1996}.

\paragraph{Morality and threat.}
In a Hobbesian society, defining ‘what is’ includes defining what is right and what is wrong. Hobbes viewed morality as subjective, and considered it as the responsibility of the Leviathan to determine what qualified as “good and bad, true and false, right and wrong”~\cite[p. 230]{williamsHobbesInternationalRelations1996}.\footnote{Hobbes' morality continues to be a topic of some interest, and debate, for many scholars, e.g.~\cite{lloydMoralityPhilosophyThomas2009}.} For Hobbes “truth is a function of logic and language”~\cite[p. 217]{williamsHobbesInternationalRelations1996} and “what is granted to that authority [i.e., the Leviathan] is the right to decide among irresolvably contested truths: to provide the authoritative criteria for what is”~\cite[p. 219]{williamsHobbesInternationalRelations1996}. This “control of normative doctrine”~\cite{lloydHobbesMoralPolitical2020} assigned to the Leviathan means that as well as defining what is right and wrong, the state can define \textit{who} is right and wrong, and of the latter, what threats they pose. If sovereignty is predicated on, or maintained using, fear, and if the sovereign has authority to determine what is to be feared, then it is in the sovereign's interest for there to exist “demons \dots{} villains''~\cite[pp. 119, 223]{neocleousCritiqueSecurity2008}, otherwise not only is the state's authority in question, as its citizens are providing obedience without receiving anything in exchange, as there is nothing to be protected from, but even its identity as a state may be threatened~\cite{heraclidesWhatWillBecome2012}. Hobbes does not refer to the benefits accruing to the state of maintaining the existence of specific threats, nor encourage their invention, however, in a criticism of religious authority, he does point out “who, that is in fear of Ghosts, will not bear great respect to those that can make the Holy Water, that drives them from him?”~\cite[p. 692]{hobbesLeviathan1985}.

\paragraph{The role of advisers.}
Hobbes discusses the value of “Counsell”~\cite[p. 303]{hobbesLeviathan1985}, distinguishing this from “Command” in that the latter “is directed to a mans [\textit{sic}] own benefit” whereas the former is “to the benefit of another man”~\cite[p. 303]{hobbesLeviathan1985}. He defines “the first condition of a good Counsellour\dots [as being that] \textit{his Ends, and Interest, be not inconsistent with the Ends and Interest of him he Counselleth}”~\cite[p. 307]{hobbesLeviathan1985} (italics in original) and describes how “the Ability of Counselling proceedeth from Experience, and long study\dots \textit{No man is presumed to be a good Counsellour, but in such Businesse, as he hath not onely been much versed in, but hath also much meditated on, and considered}”~\cite[p. 307]{hobbesLeviathan1985} (italics in original). He further adds that “The wit required for Counsel\dots is Judgement”~\cite[p. 308]{hobbesLeviathan1985} and believed that “[t]he most able Counsellours, are they that have least hope of benefit by giving evill Counsell, and most knowledge of those things that conduce to the Peace, and Defence of the Common-wealth”~\cite[p. 391]{hobbesLeviathan1985}.

\subsection{Education and discipline}
Hobbes “sought to cool men off, to pacify them, to drive them into the waiting arms of whoever might be ruling with the frightening imagery of a state of nature”~\cite[p. 88]{chapmanLeviathanWritSmall1975}. One of the mechanisms through which he intended to achieve this was through education. Hobbes had a clear view on education as being authoritarian and as being a role, indeed, a duty, of the state~\cite{bejanTeachingLeviathanThomas2010}. What Hobbes wished to be taught, according to Bejan, was Leviathan’s “‘doctrine’\dots This doctrine was no more than the existence of a ‘mutual relation between protection and obedience', which required an ‘inviolable observation'”~\cite[p. 613]{bejanTeachingLeviathanThomas2010}. Hobbes believed that the sovereign's power should be “utterly authoritarian in principle\dots and vigilantly oversee the intellectual life of his subjects from the cradle to the universities, and from there to the grave”~\cite[p. 621]{bejanTeachingLeviathanThomas2010}. 

Hobbes saw the family unit as playing a crucial role in initiating this obedience; the family is “\textit{Leviathan} writ small”~\cite[p. 77]{chapmanLeviathanWritSmall1975} (italics in original). Hobbes saw parents as “representatives of the sovereign power”~\cite[p. 620]{bejanTeachingLeviathanThomas2010}, and that “[b]y direction of the sovereign, the connection between protection and obedience is to be made quite clear”~\cite[p. 82]{chapmanLeviathanWritSmall1975}. “In teaching a child the nature of obedience in the family, a parent is teaching the nature of obedience in the state”~\cite[p. 86]{chapmanLeviathanWritSmall1975} and “[t]o teach one's children that their obedience is due when protection is given is to learn the same lesson for one's self”~\cite[p. 88]{chapmanLeviathanWritSmall1975}. The control over language that the Leviathan holds, as discussed above, underpins Hobbes' emphasis on education rather than force as the method by which the Leviathan maintained power~\cite{williamsHobbesInternationalRelations1996}, such control also helping sustain its identity~\cite{benwellDiscourseIdentity2006}. However, Bejan~\cite[p. 619, p623n17]{bejanTeachingLeviathanThomas2010} argues that Hobbes intended to stress discipline rather than education or training which are alternative translations of the \textit{disciplina} used by Hobbes in \textit{De Cive}.\footnote{In which Hobbes states “Man is made fit for Society not by Nature, but by Education”~\cite[p. 8]{hobbesCive2009}.} Hobbes did appear to consider discipline and chastisement to be productive motivators for learning, with “negative reinforcement\dots [being] an effective teacher”~\cite[p. 85]{chapmanLeviathanWritSmall1975}. Beyond educational discipline, Hobbes believed in the value of punishment, for the purpose of “correction, either of the offender, or of others by his example”~\cite[p. 389]{hobbesLeviathan1985}. He considered that “the severest Punishments are to be inflicted for those Crimes, that are of most Danger to the Publique”~\cite[p. 389]{hobbesLeviathan1985}.

\subsection{Leviathan and mini-Leviathan}
As the family was a mini-Leviathan, so too are corporations. Hobbes saw corporations as intrinsic parts of the Leviathan, even as “vital”~\cite[p. 66]{jessenStateCompanyCorporations2012} to it. However, he also identified them as potential threats, as “wormes in the entrayles of a naturall man”~\cite[p. 375]{hobbesLeviathan1985}, and in order to address these threats, they needed to be adequately governed~\cite{jessenStateCompanyCorporations2012}. This view of corporations as potential threats is similar to his view of children as potential threats to the mini-Leviathan of the family~\cite{hobbesCive2009,chapmanLeviathanWritSmall1975} if they are not adequately controlled (through education and punishment) and indeed, Hobbes uses a parent-child metaphor when referring to a form of corporation~\cite{jessenStateCompanyCorporations2012,hobbesLeviathan1985}. 

The mini-Leviathan of the corporation may pose a particular threat to the state-as-Leviathan where it is a multinational and therefore not subject to a single sovereign power~\cite{claassenHobbesMeetsModern2020,chandlerLeviathansMultinationalCorporations2005}. This can result in those corporations being able to direct and influence legislation and regulation differently in the different states they operate in~\cite{roachPrimerMultinationalCorporations2005}, frustrating attempts to achieve consistent control. Hobbes was concerned that companies would become so strong that they affected the Leviathan's own power~\cite{jessenStateCompanyCorporations2012}. Arendt points out that Hobbes could see multinational corporations as the logical endpoint of the “acquisition of wealth conceived of as a never-ending process\dots for the accumulating process must sooner or later force open all existing territorial limits”~\cite[p. 189]{arendtOriginsTotalitarianism2017}. 

Barkan argues that “corporate power and sovereign power are \textit{ontologically linked}”~\cite[p. 4]{barkanCorporateSovereigntyLaw2013} (italics in original). The “entanglement”~\cite[p. 741]{hallEntanglementsFinanceCorporate2014} between corporate businesses and states provides a link between the concept of state Leviathans and the corporation as both agent-of-Leviathan\footnote{A possibly unintended consequence, which we argue in more detail in Section~\ref{sec:cyber-and-the-leviathan}.} and as mini-Leviathan in its own right. Echoing Chapman~\cite{chapmanLeviathanWritSmall1975}, Heath \textit{et al.} refer to a corporation as “a society writ small”, but also as “an actor within the larger society in which it operates”~\cite[p. 437]{heathBusinessEthicsPolitical2010}. Part of the role that a corporation plays as agent-of-Leviathan is in the generation of “social wealth”~\cite[p. 123]{claassenHobbesMeetsModern2020} and partly through enacting regulatory control over the citizenry~\cite{barkanCorporateSovereigntyLaw2013}, even acting as a form of police~\cite{barkanRobertoEspositoPolitical2012,foucaultSecurityTerritoryPopulation2009,pasquinoTheatrumPoliticumGenealogy1991}.

\paragraph{Viability and survival.}
Hobbesian logic is based upon the avoidance of “death, pain, and disability”~\cite[p. 243]{gertHobbesReason2001} and that such “natural reason~\dots makes use of instrumental reason and verbal reason to achieve its goals”~\cite[p. 248]{gertHobbesReason2001}, that is, “the avoidance of an avoidable death”~\cite[p. 249]{gertHobbesReason2001}. Hobbes believed that “the terrour of present death” was even a valid excuse for an individual to commit a crime “because no Law can oblige a man to abandon his own preservation”~\cite[p. 345]{hobbesLeviathan1985}, and described how “since every man hath a Right to preserve himself, he must also be allowed a Right \textit{to use all the means, and do all the actions, without which He cannot Preserve himself}”~\cite[p. 5]{hobbesCive2009} (italics in original). However, such is the drive for self-preservation, that, without the governance of the Leviathan, this would lead to the ‘war of all against all’~\cite{hobbesLeviathan1985}, due to the “independence of the individuals in determining the best means to preserve their own life”~\cite[p. 74]{jessenStateCompanyCorporations2012}. As well as self-preservation of individuals, Hobbes is “unequivocal that self-preservation is the primary goal of those forming a commonwealth”~\cite[p. 115]{mcclureWarMadnessDeath2013}. Hobbes' concerns regarding survival, both of individuals and of the Leviathan itself,\footnote{Through the avoidance of war.} are echoed in discussions in classical organisational literature regarding a business's concern with ensuring its own continued viability, e.g.,~\cite{beerHeartEnterprise1979,mintzbergStructuringOrganizations1979}.

\section{Methodology}\label{sec:methods}
We collected data between October 2019 and July 2020 through 21 semi-structured interviews and by downloading each company's most recent annual report. 15 CISOs and six organisational leaders were interviewed, as shown in Appendix~\ref{sec:interviews}. The organisational leaders comprised two Chief Executive Officers (CEOs), two Chief Financial Officers (CFOs), one Non-Executive Director (NED) and one Chief Information Officer (CIO). The organisations represented a range of different industries, although with a particular weighting in one sector. As we began to notice repetition of comments from participants, we considered that data saturation may be approaching, however, data saturation is a problematic concept~\cite{oreillyUnsatisfactorySaturationCritical2013}. Ultimately we made a decision to stop gathering data on “[t]he adequacy of the sample~\dots [not] solely on the basis of the number of participants but the appropriateness of the data”~\cite[p. 195]{oreillyUnsatisfactorySaturationCritical2013}. We regularly revisited and revalidated this decision during the analysis phase, to continually confirm our judgement that the sample size was adequate. The use of annual reports as well as interview data provided triangulation, as well as the gathering of multiple perspectives through the interviewing of non-CISO participants.\footnote{Due to difficulties in obtaining access to these senior leaders, it was not possible to obtain multiple perspectives from every organisation, as had been our original intention.}

One of the researchers is a practising CISO and used their own network of professional contacts to recruit participants, effectively producing a ``snowball sample''~\cite[p. 135]{hammersleyEthnographyPrinciplesPractice1995}. Access to Board members is difficult outside of a professional environment and, therefore, these participants were approached through CISOs who were participating in the research, as well as our personal networks. Participants were not compensated for their participation in this research. Interviews took place either face-to-face at their own office locations or online, the latter in response to the COVID-19 pandemic which commenced during data collection. We recorded the interviews and transcribed them as soon as possible following each interview, capturing non-recorded aspects such as body language and spatial information, in a handwritten journal immediately following each interview. Interview guides were prepared with prompts to be used as necessary, however, interviews were approached as conversations rather than extractions of data and, therefore, these were not used in a strict manner, following Hermanowicz~\cite{hermanowiczGreatInterview252002}.

The majority of Participants, and all of those in CISO roles, self-identified as male. In addition, there was limited ethnic diversity in the study and all participants can also be considered to be `elites'. This lack of diversity, however, reflects a broader lack of diversity in the cyber-security industry. We received approval from our our institution's Research Ethics Committee for self-certification before beginning the research and designed the study to minimise both the collection of personally identifiable information and the risk of indirect identification. Participants were provided with consent forms and information sheets two working days before each interview which explained how data would be anonymised and protected. Participants were anonymised, with randomly assigned pseudonyms being utilised which, in this paper, have been substituted for participant numbers. We redacted any sensitive or potentially identifiable information during transcription and destroyed all recordings following transcription. 

Interview transcripts were analysed inductively and coded in multiple cycles using NVivo 12~\cite{QualitativeDataAnalysis} and applying a variety of coding types, following~\cite{saldanaCodingManualQualitative2016}. Subsequently, we applied a deductive approach in order to categorise and rationalise the codes. A similar method was used to analyse annual reports, however, as this was performed subsequent to the interview coding cycles, coding became more deductive, as codes and concepts determined at the previous inductive stage were, consciously and unconsciously, reused. We developed themes from our data following~\cite{braunUsingThematicAnalysis2006,braunReflectingReflexiveThematic2019}. Following others, e.g.~\cite{saldanaCodingManualQualitative2016}, we produced analytic memos throughout as well as using diagrams to explore relationships and to identify and explore themes developed from the data, again following~\cite{braunUsingThematicAnalysis2006} and combining several methods from Salda\~{n}a~\cite{saldanaCodingManualQualitative2016}.

As has been established within cyber-security scholarship, e.g.~\cite{burdonSignificanceSecuringCritical2019}, qualitative research that follows an interpretive paradigm is an effective means of studying cyber-security practice. The use of semi-structured interviews to gather data is also well established, e.g.~\cite{singhInformationSecurityManagement2013,ashendenCISOsOrganisationalCulture2013,mooreIdentifyingHowFirms2015,burdonSignificanceSecuringCritical2019}, as is analysis of annual reports to derive insight about businesses~\cite{joshiExplainingITGovernance2018,zmudSystematicDifferencesFirm2010}, including the use of document coding~\cite{zmudSystematicDifferencesFirm2010} as discussed further below.

\section{Findings}\label{sec:findings} 
Our thematic analysis produced multiple themes, a number of which had Hobbesian connotations and we present these below.\footnote{Other themes will be explored in future research.} Section \ref{sec:microcosm} shows these organisations as mini-Leviathans, Section \ref{sec:viability} covers aspects of survival articulated by these businesses and Section \ref{sec:org-society} sets out these organisations' role in wider society.\footnote{Throughout the remainder of the paper, we use ‘business’ and ‘organisation’ interchangeably for ease of reading. For the avoidance of doubt, all of the organisations referred to are commercial businesses.} Certain quotations are not attributed to limit the risk of identification. No direct quotations from annual reports are included for the same reason and where single quotation marks are used, these indicate paraphrasing. 

\subsection{The organisation as Leviathan-writ-small}\label{sec:microcosm}
The organisations in this study demonstrated a number of features that indicated their operation as miniature societies. These included references to the organisation having both a culture and ‘values’, which were observed in the majority of annual reports and approximately half of the interview transcripts. The culture of the organisation affected how the participants viewed their ability to influence cyber-security outcomes. This included a continuum regarding risk posture being described by many participants, from “ludicrously conservative” to less conservative (CISO8). These organisations regulated employee behaviour, indicated by specific mentions of mandatory standards of behaviour and conduct, on which staff were trained, measured and, in some cases, penalised for non-compliance. Organisations in this study expected their staff to comply with their policies, adhere to their ‘core values’ and even to adjust their ‘mindset’ as part of working for the company. CISOs described having responsibilities to “introduce the right sort of behaviours and judgements in our workforce” (CISO2). 

Cyber security was an area of both discipline and punishment. Part of the role of the cyber-security function in these organisations was “to hold feet to the fire” (CISO5). Compliance with security policies and standards was mandated and failure to comply could result in “[being] on a disciplinary” (CISO3), with staff facing “disciplinary action~\dots even if they’ve done nothing wrong” (CISO12). Despite a number of CISOs being keen to avoid the characterisation, cyber-security teams were seen as performing a policing function. This included specific references to being “the police” as well as more subtle references to “stop[ping] people having fun” (CISO1) and “trying to find [staff] doing wrong” (CISO8). 

As well as disciplinary action, the organisations in this study linked staff remuneration, particularly at a senior level, to cyber security through their performance objectives, both explicitly and implicitly. These included multiple references to punishment of staff through “clawbacks” of bonus payments in particular. Triggers for the latter included reputational impact, direct losses, regulatory investigations, contractual breaches and, commonly, general failures in risk management. More broadly, remuneration was dependent on a number of factors including both risk management and ethical performance, areas that were commonly linked with cyber security. In some cases, cyber-security objectives were described as specific measures relating to bonus payments, as were measures relating to the completion of mandatory compliance training. 

Cyber security was also associated with state punishments, whether through fines, “other sanctions~\dots from government” (CFO2) and even incarceration. Some of those punishments were viewed as useful by participants. CISO9 described how “it was only when the likes of Marriott Hotels or BA [British Airways] started to get massive fines relating to personal data that suddenly Boards sat up and took notice”. The annual reports also indicated that these organisations were concerned that cyber-security failings would lead to punishments, either from regulators or through legal action, with explicit references to enforcement and censure that were considered to be threats to organisational viability, as discussed in Section~\ref{sec:viability}.

\subsubsection{Cyber security as pedagogy}\label{sec:cyber-pedagogy}
As well as applying discipline, a key role of the CISOs in our study was educating staff. References to cyber-security education were made by multiple participants as well as across the majority of annual reports. This involved not just “making sure they’re [i.e., staff] educated well” (CISO3) but also “[making] cyber-security meaningful for them…on a personal level” (CISO11). There was a need to “educate” because cyber security was “another language” (CISO5).
Senior stakeholders were also included, and their education was specifically called out in a number of annual reports, particularly the recency of such education. 

Various methods used by these organisations to educate staff and stakeholders on cyber security were mentioned, including “visual breakdowns” (CISO2) and “games” (CISO11), as well as testing of staff, particularly through simulated phishing attacks. Many of the references in the annual reports to this education included the modifier ‘mandatory’. There were indications of the deliberate use of fear in relation to cyber security, such as the use of “war games [with senior leaders]~\dots and you watch them shit themselves” (CISO11). CEO1 described the value in using fear, stating that “[when staff] see the art of the possible and it's scary~\dots they say okay, I'm gonna whine less”. CISOs acknowledged that “it’s very difficult for a conversation [about cyber security] not to gravitate back to being scary and inevitable” (CISO8) but were conscious of the risk of “scaremongering” (CISO5).

Cyber security was consistently characterised by participants as having an ethical or moral dimension, with ‘rights and wrongs’ relating to cyber security being a common refrain observed throughout the data. One CISO described their department as “the moral police force of the company” (CISO8). Cyber security personnel had a “duty to communicate risk” (CFO2) and to “hold [the organisation] to account to make sure they're doing the right thing” (CISO7). Cyber-security failures at a single organisation could have wider societal impacts; one participant described their company as “the soft underbelly” for their customers, who themselves supported wider societal goals such as distribution of food. The articulation of cyber threats in moral terms was also consistent. This included references to cyber threat actors as “bad guys” (CISO9) and a statement that “the mission [of the cyber-security function] really is to protect against crime” (CISO8).

\subsubsection{The CISO as advisor}\label{sec:different-roles}
As well as being an educator, the CISO for these organisations also performed a role as a form of advisor to the organisation. This was summed up by one senior leader who described the need to be told “no you don’t need to be worried about that, yes you do need to be worried about this” (CEO1). Many CISOs articulated this role explictly, such as being “a trusted advisor to the business~\dots to provide guidance, provide advice” (CISO12). Annual reports also indicated the advisory role that specialist risk management functions, including cyber security, provided to these businesses, describing the use of such advisers in providing both predictability and interpretation of uncertainty. Such advisers were trusted to provide “judgements” (CISO2).

CISOs were aware that their functions were “going to be there for the long term that’s for sure” (CISO3), with senior leaders agreeing that it was “certainly not gonna get less important” (CFO1). However, CISOs in this study indicated concerns that they could be subject to punishment through job losses. They were “not under any illusions [as] to where accountability sits” (CISO3). They knew that they “wouldn’t escape the spotlight” (CISO2) and “that it’s implicit with our role, if something goes wrong, you’re the guy [that gets fired]” (CISO12).

\subsection{Viability and survival}\label{sec:viability}
The large majority of businesses in our study expressed cyber-security threat as a survival-level concern, with cyber-security incidents being able to “destroy the business” (CEO1) or “bring the company to its knees [and] drive us to bankruptcy” (NED1). In many cases cyber security was explicitly referenced in the viability statements made in their annual reports.\footnote{All organisations in this study made reference to various threats to their ongoing viability in their annual reports, however, this is unsurprising given that such statements are a requirement of the UK Corporate Governance Code~\cite{financialreportingcouncilUKCorporateGovernance2018}.} Cyber-security incidents were positioned as threats to this viability by many of these organisations, with fear of regulatory action and associated fines and reputational damage being a prime concern. Such incidents were considered as existential threats and phrases such as “absolute catastrophe” (CISO11) and “disastrous” (CISO9) were common. Cyber security functions were, in a number of cases, seen as assuagement against this threat to viability, with CISOs providing “a level of comfort” (CISO14) to their senior leaders.

Cyber security threats were considered to be both permanent and fearful. They were “really scary” (NED1) and resulted in “sleepless nights” (CISO1). For these participants, it was “when not if [a cyber-security incident occurs]” (CISO14), and they needed to “accept the fact that we are going to be compromised” (CISO11). This normalisation of cyber-security incidents was also indicated by senior leaders, with them being characterised as “the kind of things that happen all the time” (CEO2). NED1 described how cyber security was “gonna get worse not better”, with CFO1 describing cyber security as “a continuing moving goal post”. Cyber-security threats were “so sophisticated [and] change almost on a daily basis” (CISO14), and there were “troubled times ahead” (CISO4). Similar statements were observed in the annual reports, with references to the ‘sophisticated’ and ‘continual’ nature of cyber threat being common throughout.

As well as explicit references to the potentially catastrophic nature of a cyber-security incident, we also observed more implicit references to fear in connection with cyber security, such as mentions of cyber crime and cyber terrorism. Threats were also seen to originate from other sources, including hacktivists and, in particular, nation states. One CISO described how “the whole concept of state sponsored threat actors is frightening” (CISO14). Some of these states were named, e.g., “the Chinese~\dots somebody sitting in Siberia~\dots North Koreans” (CEO1), “Iran~\dots China~\dots Russia” (CISO8), whereas other references were generic, unnamed nation states. Explicit references to cyber warfare were also observed, with cyber security being a “method of attack against the nation” (CISO3). Cyber security was positioned as a component of national security by multiple participants and annual reports, highlighting their organisation's role within this, which we describe further below. Metaphors of war in connection with cyber security were common throughout the data, including “attack”, “defend”, “war stories” and “war games”.\footnote{Other stereotypically masculine language and concepts were observed throughout much of the data. These included participants referring to their businesses being like a “bearpit” (CISO6), needing to avoid being “too soft” (CISO13) and being “browbeaten” (CISO1). Similar language was seen in the annual reports, with the use of conventionally masculine concepts such as aggression, conflict, strength and even penetration being common. Both the interview data and the annual reports also featured multiple references to competition, another masculine archetype.}

\subsection{The organisation in wider society}\label{sec:org-society}
Virtually all of the organisations in this study articulated, through their annual reports, the broader societal role that they played, and the benefits that society derived as a result. These included the contributions those companies made through investment, through community support and the delivery of ‘critical services’ to that society. Annual reports included language relating to societal obligations and responsibilities, societal impacts, and even direct references to the social contract. Participants also mentioned the role that their organisations played in wider society. Some referred to their organisations as being critical to the functioning of the UK economy, while many referred to the role that they played in national security. One organisation's cyber-security department was assessed by a UK military agency before they were “allowed to bid” on a contract, with “the quality of the cyber-security team [being] very much the litmus test”. Another organisation's CISO described being “conscious of~\dots our responsibility~\dots to defend against [a cyber attack on the country]” while one of the CEOs described being obligated by government to take actions in relation to national security. Other references to the role these organisations played in national cyber security included participating in national security working groups and being regularly assessed by security services and other government departments.\footnote{One participant described having recently been visited by representatives from the UK intelligence services directly preceding our interview for this research, which may have affected their responses.} It appeared there were double standards regarding evaluation by government, with government departments charged with assessing these businesses responding with “oh Lord no, we would never achieve it” when asked if they complied with the same requirements. There were also references to to invitation-only and industry-specific information exchanges with representatives from state security services where specific threats were shared with attendees as well as indications of more indirect governmental influence. The latter included senior leaders being invited by government departments to participate in “roundtable discussions” and being encouraged to utilise certain frameworks. One CEO described how “the UK government has been quite vocal [on cyber security]”, with another CEO stating that “[governments, plural] keep an eye out, which works in ways that neither you or I need to know how it works but they keep an eye out”. A number of annual reports alluded to potential negative impacts on revenue if the focus of their government customers moved away from security-related products and services. It was also noted that a number of these organisations had senior leaders who either currently or previously held positions within the military, government or quasi-government organisations.

\subsubsection{Security versus freedom}\label{sec:security-freedom}
A perceived dualism between security and freedom in relation to cyber security was indicated within the data. CISO4 described there being “an amount of disruption that is necessary in order to do the right thing”. Security controls were “very tight” (CFO2) with an expectation that they would “get tighter and tighter” (NED1). CEO1 linked this to fear, describing how “the more that we go and scare ourselves~\dots the more the organization becomes willing to tolerate some inconvenience in what it does”. 

Organisations in this study surveilled their staff in a number of ways. These included monitoring of company vehicle use, for both health and safety and ethical reasons, i.e., vehicle emissions, as well as monitoring of technology systems for cyber security and IT reasons. As well as surveilling their staff, there were also examples of organisations surveilling their customers in terms of how their products and services were used by them, although with an acknowledgement from CISO8 that “it is a tough balance, not everybody wants it [i.e., monitoring of product usage]~\dots some people are really paranoid”. They described difficulties in achieving “a balance between inspection and surveillance”, and potential impacts upon “free speech”, with their customers holding different views ranging from “absolutely no problem [with monitoring]” to monitoring of activity being “abhorrent”. There were also indications of deference, even servility, to wider surveillance occurring at state level, as per the final CEO comment in Section~\ref{sec:org-society}.

\section{Discussion}\label{sec:discussion}
In this section, we unpack our findings by applying a number of Hobbesian concepts in an attempt to provide deeper meaning. We first describe in Section~\ref{sec:perpetuall-feare} how Hobbes can be used to read the threat to survival that cyber security posed to these organisations. Next, in Section~\ref{sec:cyber-obedience} we apply Hobbes to cyber-security related discipline and punishment enacted by these organisations. Finally, in Section~\ref{sec:cyber-and-the-leviathan} we explore the wider role of cyber security in the context of the state Leviathan.

\subsection{“Perpetuall feare”}\label{sec:perpetuall-feare}
Cyber security is often characterised as fearful, with cyber threats being both permanent and evolving~\cite{ehrlicherCouncilPostEvolution2021}, and cyber attacks being seen as inevitable~\cite{pearlsonCyberattacksAreInevitable2021}. The businesses in our study seemed to agree, with cyber threat appearing to be normalised. Participants, both CISO and non-CISO, considered these threats to be enduring and businesses needed to accept that they would be compromised. This implies a “perpetuall cyber warre”~\cite[p. 120]{stevensCyberSecurityPolitics2016}. These businesses existed in “continuall feare”~\cite[p. 186]{hobbesLeviathan1985}, threatened by “death, poverty, or other calamity”~\cite[p. 169]{hobbesLeviathan1985} arising from something, i.e, cyber security, that was not well understood, even mystical,\footnote{The impacts of cyber-security threats may be considered by these organisations as ‘real’ but the threats themselves, including their sources, may be considered more ephemeral. The mystical nature of cyber security was a separate theme developed from the data which will be explored in future research.} as “perpetuall feare, [is] always accompanying mankind in the ignorance of causes”~\cite[p. 169-70]{hobbesLeviathan1985}.\footnote{The state of nature may even be considered as a “secular hell”~\cite[p. 618]{bejanTeachingLeviathanThomas2010}. Similar metaphors have been used with reference to cyber security, such as “cyber hell”~\cite[p. 480]{shrobeConclusion2018} and “cyber apocalypse”~\cite[p. 105]{stevensCyberSecurityPolitics2016}.} The role of the CISO may be valued as one “that can make the Holy Water”~\cite[p. 692]{hobbesLeviathan1985} that provides protection from fearful things, and, therefore, is motivated to maintain the fear and dread that underpins their value.\footnote{The potential that this offers for ‘cyber sophistry’ will be explored in future research.}

Survival was clearly a concern for these businesses. Cyber security was positioned as a threat to viability by many of these organisations, with fear of regulatory action and associated fines and reputational damage in particular being a prime concern. Such concerns with viability and survival, arguably the primary motivation for businesses~\cite{beerHeartEnterprise1979}, are analogous with seeking to avoid punishment that could lead to “pain, and disability” and, ultimately, “death”~\cite[p. 243]{gertHobbesReason2001}. The punishments they sought to avoid were enacted by the larger Leviathan of the state and, as mini-Leviathans, these businesses cascaded this concept, instituting their own mechanisms of punishment for their employees, as we discuss below. Internal experts were positioned by many of these organisations as ‘guards’ that protected them against the various harms that they faced. Cyber security functions were seen as assuagement against threats to business viability and helped these businesses to manage the uncertainty they experienced as a result of these threats and ensure their continued survival. This allowed them to shape their future to a certain extent, aligning with Hobbes’ encouragements towards continued attentiveness to threats~\cite{stevensCyberSecurityPolitics2016}, but also provided a resource that articulated and predicted those threats, based on both past, and imagined future, events.

\subsubsection{Permanent cyber emergency}
Many of these organisations played a role in national cyber security, including participating in invitation-only national security working groups. Such fora, in which government intelligence services share details of cyber threats with specific industries, demonstrate the role that governments play in maintaining a state of permanent cyber emergency. They provide a mechanism through which governments can both maintain fear and amplify it. This could be achieved through exaggeration or even fabrication, particularly when considering the reliance of the state Leviathan on the persistence of this fear.\footnote{The use of falsehoods in the service of continued peace was something that Hobbes appeared to support~\cite[p. 703]{hobbesLeviathan1985}~\cite[p. 224, pp. 290-1]{arendtFuture2006}, although, as Bejan summarises, this reading is debated and other scholars have found his intentions “far less sinister”~\cite[p. 623n13]{bejanTeachingLeviathanThomas2010}. However, Hobbes was clear that the authority of the sovereign was absolute, even in matters of “Prophecy”~\cite[p. 466-469]{hobbesLeviathan1985}.}

A permanent emergency offers benefits as a “master narrative”~\cite[p. 18]{smithContrastingPerspectivesNarrating2008} that can be invoked to support actions taken by businesses and individuals within that business, whether to justify investment or to justify restrictive controls such as surveillance, as we discuss in Section~\ref{sec:cyber-and-the-leviathan}. The positioning of cyber security as warfare, which is a narrative repeated by both media and governments~\cite{stevensCyberSecurityPolitics2016}, establishes that concept in the minds of all parties to that war, whether attacker or attacked. Adversaries, or even just those who disagree, will respond to the narrative of cyber security-as-war and then treat it as such, focusing on attack and defence, rather than seeing it as anything else, for example, as a collective problem of identifying and addressing weaknesses that threaten all. This could lead to actions that have unintended consequences, such as state purchasing, and hoarding, of vulnerabilities, e.g. ~\cite{hoeksmaNHSCyberattackMay2017}. Cyber security-as-collective-problem could be considered a “flattened narrative”~\cite{farleyAmusingMonsters2001}, with preference instead provided to the cyber security-as-war concept which supports the maintenance of existing power structures. This can also be considered as indexing a “meta-narrative”~\cite[p. 3]{Gillian2014IWaB} of existing or ‘traditional’ enemies~\cite{stevensCyberSecurityPolitics2016}, and the “superiority of~\dots the West”~\cite[p. 172]{neocleousCritiqueSecurity2008}, as suggested by the references to (ex-)communist states\footnote{Who are also competing state Leviathans.} observed in the data. There is an almost paradoxical relevance of both Foucault's and von Clausewitz's perspectives on war, with cyber war being a “mere continuation of policy by other means”~\cite[p. 12]{vonclausewitzWar1873} and cyber politics being “the continuation of war by other means”~\cite[p. 15]{foucaultSocietyMustBe2004}.\footnote{It is outside the scope of this paper to explore these arguments in more detail, however, future research will apply a number of Foucauldian lenses to our Findings.} Each of these also provides economic benefits~\cite{neocleousCritiqueSecurity2008}, as we discuss below.

Positioning cyber security as an existential threat, as a war with “apocalyptic”~\cite[p 121]{stevensCyberSecurityPolitics2016} consequences, may also allow for exceptionalism and deviance from existing laws, both national and supranational. Hobbes explicitly permitted defiance of law if motivated “by the terrour of present death”~\cite[p. 345]{hobbesLeviathan1985}. Such exceptionalism based on existential threat can be observed in modern societies, e.g.~\cite{USAPATRIOTAct2001,ukgovernmentRegulationInvestigatoryPowers2000,ukgovernmentTerrorismAct20062006}, including in relation to cyber security threats~\cite{walkerCyberTerrorismLegalPrinciple2006}. References to national security also connect with this warfare motif. As the nation is ‘under threat’ then there is a collective sense of conflict and, therefore, a suggestion that everyone has to play their part.

The fear generated by these threats propels citizens into “the waiting arms of whoever might be ruling”~\cite[p. 88]{chapmanLeviathanWritSmall1975}. Such fear may be a “necessity-justification”~\cite[p. 89]{chapmanLeviathanWritSmall1975} for enduring power, and, as Chapman suggests, it is straightforward to conceive of such “justification~\dots [occurring] at the state level, as a function of real or manufactured inter-state crises”~\cite[p. 89]{chapmanLeviathanWritSmall1975}, particularly with regard to threats that are hard to understand or somewhat ephemeral in nature, such as those related to cyber security. If cyber-security threats result in fearful and bewildered citizens, those citizens are easier to moderate. Educating citizens on, or even communicating the existence of, cyber threats may couple “paranoia with pacification”~\cite[p. 90]{chapmanLeviathanWritSmall1975}.

Businesses that publicly articulate their cyber-security capability, through references to the existence of dedicated personnel and the actions they are taking to mitigate cyber risk, are demonstrating their strength and their readiness for war in a  “calculated presentation”~\cite[p. 92]{foucaultSocietyMustBe2004}. Such pronouncements, particularly in annual reports, also serve to maintain the organisation's power by “memorializ[ing]” what the organisation has achieved, arguably also creating “an obligation”~\cite[p. 67]{foucaultSocietyMustBe2004} for future leaders of that organisation. 

\subsubsection{Cyber discourse}
The collocation of certain words observed in the data, e.g. ‘sophisticated’ and ‘threat’, which are also seen in broader cyber-security discourse, e.g.,~\cite{noonanBankEnglandbackedCyber2021}, may carry “encoded ideologies”~\cite[p. 113]{benwellDiscourseIdentity2006} that also serve to maintain power structures. References to ‘nation state’ alongside ‘cyber threat’ carry an association of war being waged, particularly by previously established ‘enemies of the West’. As “packaged, homogenized violence”~\cite[p. 160]{baudrillardConsumerSocietyMyths1998}, such references not only maintain hegemonical power (‘we’ are threatened by ‘them’ therefore we must take action) but also provides a means by which citizens are mollified, arguably even tranqullised, as well as driving consumption~\cite{baudrillardConsumerSocietyMyths1998}, as we discuss in Section~\ref{sec:cyber-and-the-leviathan}.

Hobbes saw the importance of the Leviathan having control over language. By maintaining discourse that defines, or repeats, who and what are threats, and the relative urgency of those threats, the state can maintain broader narratives of fear, war, friend and enemy, good and bad, and right and wrong. Such narratives in connection with cyber security featured in our data but, in particular, the articulation of cyber threats in moral terms was consistent. A moral association may strengthen the power and importance of these threats for citizens but also result in unquestioning acceptance of those positions. Although morality may (arguably) be subjective~\cite{zimmermanMoralObligationObjective2006},\footnote{It is outside the scope of this paper to explore the ongoing and unresolved philosophical debate concerning this highly contentious position.} it may be \textit{experienced} objectively in everyday life~\cite{hofmannMoralityEverydayLife2014} and, therefore, by assigning a moral dimension to cyber security, citizens may be discouraged from challenging the ‘need’ for intrusive controls associated with it.

The use of specialist cyber-security language, which is inaccessible to non-specialists, provides power to those that can understand it, and this power is increased when there is an interpretation being provided. An interpretation provides an opportunity, conscious or unconscious, to imbue its translation with other meanings, whether moral, political or emotional. Language is a means by which reality is both experienced and constructed~\cite{benwellDiscourseIdentity2006}, with those who have the power to interpret specialist or ‘foreign’ language also having the power to construct reality for their audience. Cyber security may offer a channel through which sentiments and beliefs that are beneficial to the Leviathan can be established and maintained, such as those relating to ‘enemy threats’ or those relating to ‘security versus privacy’, a questionable dualism~\cite{neocleousCritiqueSecurity2008} that we discuss below..

Cyber security in both academic and mass media communication abounds with military tropes, e.g.~\cite{kanniainenCyberTechnologyArms2019,limnellCyberArmsRace2016,coreraNHSCyberattackCame2017,bondBritainPreparingLaunch2018} and many metaphors of war were observed throughout our data. Such militaristic references may be motivated by a desire for those who work in cyber security, most of whom are male~\cite{peacockGenderInequalityCybersecurity2017},\footnote{Modern corporations are also male-dominated~\cite{connellGlobalizationBusinessMasculinities2005}, often with hierarchical structures that feature inherent masculinity through militaristic associations~\cite{carverMenMasculinitiesInternational2014}.} to cast themselves as heroic, a masculine trait that is strongly Hobbesian~\cite{distefanoMasculinityIdeologyPolitical1983}. The perception of always being ‘at war’ from a cyber-security perspective, besides the ontological ‘comfort’ this may provide~\cite{mitzenOntologicalSecurityWorld2006}, may also be motivated by a masculine desire for such valorous narratives, demonstrating masculinity through “metaphor, and bravado”~\cite[p. 115]{carverMenMasculinitiesInternational2014}. Such language may also be deeply performative~\cite{carverMenMasculinitiesInternational2014,butlerGenderTroubleTenth2002}. Cyber security professionals may possess a distinctive and exceptional “power” that helps form their heroic identity, namely “knowledge” and “right method”~\cite[p. 642]{distefanoMasculinityIdeologyPolitical1983} which represents “the requisite special weapon of the epic hero”~\cite[p. 642]{distefanoMasculinityIdeologyPolitical1983}, and, similar to Hobbes' self-conception as heroic, cyber-security professionals may be “proposing a solution to a predicament that [is] more masculine than human in tenor”~\cite[p. 643]{distefanoMasculinityIdeologyPolitical1983}.

\subsection{Protection in exchange for (cyber)~obedience}\label{sec:cyber-obedience}

Cyber security was an area of discipline and of punishment. Organisations in this study required obedience from their staff, in terms of policy compliance, as well as alignment with standards of behaviour. Obedience, whether through completion of mandatory training, compliance with cyber security policies and standards, or through effective management of cyber-security risk, was mandatory and non-compliance would be punished. As well as disciplinary action, non-compliance resulted in impacts upon staff remuneration, particularly at a senior level. The potential for the organisation to remove a level of security from its staff, in terms of the security of continued employment and income, suggests what Barkan\footnote{Discussing Esposito~\cite{espositoBiosBiopoliticsPhilosophy2008}.} describes as an “immunitary dynamic”~\cite[p. 89]{barkanRobertoEspositoPolitical2012}. As with the Leviathan, these businesses were providing protection in exchange for obedience and if this obedience was not received, their protection could be removed. 

As discussed above, the Leviathan is tyrannical. While our study has not focused on all aspects of corporate life, some indications of tyranny that align with Friedrich and Brzezinski’s~\cite{friedrichTotalitarianDictatorshipAutocracy1961} description of typical totalitarian features were observed. The cyber security departments in these organisations appeared to function as an official police force, despite CISOs wishing to avoid this characterisation, and performed surveillance of staff. They acted as agents of the mini-Leviathan, applying discipline and punishment. Beyond cyber security, these organisations applied punishments if staff did not comply with their dictates, including if they behaved contrary to their values. This may also have been motivated by a lack of parity between the interests of the organisation and the interests of individual employees. Punishment was imbued with morality by extending a concept of ‘doing the right thing’. These organisations educated and indoctrinated their staff, with fear being a component of these processes. 

Cyber security was also associated with state-directed punishment. The threat of punishments relating to cyber security had a regulatory effect on these organisations, with considerable attention paid in the annual reports to addressing how compliance was monitored and enforced, including references to the organisational capabilities charged with these responsibilities. Organisations in this study wanted to ‘do the right thing’ in order to avoid punishment by the Leviathan and internally, as mini-Leviathans, instituted mechanisms of punishment for their employees, extending that same concept. The references that participants made to the ‘usefulness’ of cyber security incidents affecting other organisations, including explicit references to those that resulted in regulatory action, suggest a Hobbesian view of punishment as providing examples for others but also a spectacular nature to cyber-security punishment. This is similar to that described by both Foucault~\cite{foucaultDisciplinePunish1991} and Farley~\cite{farleyAmusingMonsters2001},\footnote{Farley also invokes Hobbes in his exploration of state punishment.} and punishment itself is a representation of power~\cite{foucaultDisciplinePunish1991}. The Leviathan is terrorised by cyber-security threats, whether real or imagined, which, in the most fearful type, arise from the Leviathan's known enemies. It expects its “lesser Common-wealths”~\cite[p. 375]{hobbesLeviathan1985} to take action against these threats for the benefit of the larger commonwealth. Failure to obey results in punishment by the Leviathan, such punishments being public spectacles that provide examples to others. 

\subsubsection{CISO as teacher and “counsellour”}
The CISOs in these organisations were educators, which included “teaching~\dots obedience”~\cite[p. 86]{chapmanLeviathanWritSmall1975} and applying discipline. They taught staff about the existence of cyber-security threats, communicated a defined set of rules, indoctrinated them into acceptable behaviours, monitored their compliance against these, and punished them when they transgressed. Staff were educated that both they, and the organisation itself, were subject to cyber-security threats. In order for the organisation to mitigate those threats, protecting both itself and its staff, those staff must forgo certain liberties and agree to be regulated. CISOs utilised fear in their instruction, aligning with Hobbes' template~\cite{chapmanLeviathanWritSmall1975}. 

As well as being teachers, CISOs were advisors, “Counsellours”~\cite[p. 391]{hobbesLeviathan1985}, for these businesses. They had “knowledge of those things that conduce to the Peace, and Defence of the Common-wealth”~\cite[p. 391]{hobbesLeviathan1985}. Although such counsellours may be expected to have consistent “Ends and Interest” with the organisation, they may still derive “benefit by giving evill Counsell”~\cite[p. 391]{hobbesLeviathan1985}, particularly if that benefit is continued employment. One area where the CISO may not share equivalence with Hobbes' counsellours is in their risk of scapegoating. Hobbes' view was that “he that demandeth Counsell, is Author of it; and therefore cannot punish it”~\cite[p. 304]{hobbesLeviathan1985}, however, the CISOs in this study indicated concerns that they were subject to punishment through job losses.

\subsection{Cyber security and the Leviathan(s)}\label{sec:cyber-and-the-leviathan}
It is not in the interests of a Hobbesian society to achieve “complete security”~\cite[p. 184]{arendtOriginsTotalitarianism2017}. Both the relative novelty of cyber-security threats and the continued emergence of new types of such threats, including the ‘sophisticated’ aspects thereof, can be viewed as “new props from the outside”~\cite[p. 184]{arendtOriginsTotalitarianism2017} that stoke the flames of the possibility of war, particularly when attributed to nation states. These threats also offer “new and ever-growing fields for the honorable and profitable employment”~\cite[p. 28]{hobsonCapitalismImperialismSouth1900} of citizens,\footnote{Specifically, the employment of “sons”~\cite[p. 28]{hobsonCapitalismImperialismSouth1900}.} particularly the bourgeoisie who are appeased by new job opportunities, and further stimulate consumption and growth~\cite{arendtOriginsTotalitarianism2017}.

In a (post)modern world where threats to the state are less obvious or apparent, i.e., there is no obvious invader on the doorstep, particularly since the end of the Cold War, the inclination of the citizen towards obedience may be weaker. The state may, therefore, feel the need to motivate obedience by making it clear that it is still offering protection, but not against obvious invaders. Rather, it is against opaque, mysterious, and highly sophisticated threats, from which the state is providing protection. Not only do these threats need to be explained by specialists, due to their complexity, they also need to be ‘sold’ to citizens through education. It is even conceivable that such teachings could be contrary to “true Philosophy”~\cite[p. 703]{hobbesLeviathan1985} but serve the benefit of the state, as well as securing the continued employment of the teacher~\cite{arendtFuture2006}. This could motivate the embellishment of any threats communicated. It may be more advantageous for the state Leviathan to have such education delivered not through a state organ but rather through another component of society such as businesses. Rather than a conscious decision taken by the Leviathan this may be a fortuitous benefit, but one that it seeks to encourage through, e.g., communicating the ‘responsibility’ that businesses have in protecting wider society against cyber threats~\cite{ukgovernmentNationalCyberSecurity2016}.

Dedicated cyber-security functions support, and repeat, messages relating to a broader security agenda~\cite{neocleousCritiqueSecurity2008}, both among a company’s employees and their customers. There is a wider security industry that “must…ensure that security is never really achieved”~\cite[p. 156]{neocleousCritiqueSecurity2008}. This provides commercial benefits, as well as supporting an insecurity that is relied on by the state to achieve its aims~\cite{neocleousCritiqueSecurity2008}, as previously noted by Arendt~\cite{arendtOriginsTotalitarianism2017}. If states ultimately seek the perpetuation of (at least partial) insecurity, then it may be in their interests to define ‘security’ in an insecure manner, at the same time encouraging organisations and wider society to achieve a level of ‘insecure security’. Recent attempts by governments to weaken or circumvent strong encryption, e.g.~\cite{thomsonLowBarrDon2019,GoogleAppleCriticise2019}, some successful, e.g.~\cite{taylorAustraliaAntiencryptionLaws2019}, can be argued as demonstrating this desire.\footnote{Although increased surveillance may be possible without weakening encryption~\cite{gasserDonPanicMaking2016}.}  In addition, motivating organisations to operate a cyber-security function that inures employees, who are also citizens, to increased and intrusive surveillance and monitoring may also contribute to this same ‘insecure security’, albeit potentially providing associated benefits to those employees, such as greater privacy.

These organisations taught their staff about the existence of cyber-security threats, communicated a defined set of rules, indoctrinated them into acceptable behaviours, monitored their compliance against these, and punished them when they transgressed. In order for both the organisation and their staff to be protected from these threats, staff must forgo certain liberties and agree to certain controls, such as being surveilled. In this manner, the employee-as-citizen is indoctrinated into a mindset of being subject to cyber threats and becomes inured to the custom of exchanging privacy for security. Considering power relations as interactive, processual and two-way~\cite[p. 89]{benwellDiscourseIdentity2006}, cyber security within business can be seen as a mechanism through which citizen-employees participate in the maintenance of hegemonic power.\footnote{Other hegemonical linkages included the roles that these businesses played in national security and the presence of senior leaders representing military or governmental actors, as well as references to both direct and indirect governmental influences on these organisations.} Cyber security personnel ensure that messages of insecurity and threat are repeated, as well as performing a policing role that normalises surveillance, while other citizen-employees support hegemonic power in the role they play as the surveilled.

Cyber security can be used to terrorise citizens into compliance and to justify their surveillance. Cyber-security controls have an effect not just on the employee-as-employee but also the employee-as-citizen. Educating employees on what they need to be protected from, and what they need to obey in order to be protected, may, directly or indirectly, inure or condition employees towards broader obedience, including acceptance of controls that could be used beyond purposes of cyber security, providing benefits to the state beyond citizen protection.\footnote{“The most potent weapon in the hands of the oppressor is the mind of the oppressed”~\cite[p. 137]{bikoBlackConsciousnessQuest1981}.}  The use, and acceptance, of surveillance in these organisations may function as a normative control~\cite{beechNatureDialogicIdentity2008}, conditioning or preparing staff (citizens) to be surveilled in wider society and supporting a wider meta-narrative in relation to security versus privacy. The Leviathan-writ-small of the business plays a role on behalf of the state Leviathan in conditioning the employee-as-citizen towards obedience, and surrendering of liberties, in exchange for protection from threat. The opaque and relatively unseen nature of the threat, which requires specialists to deliver education about its existence, is beneficial to the state for ensuring continued obedience, and even in maintaining its own identity~\cite{heraclidesWhatWillBecome2012}, and its own history~\cite{simsYouBastardNarrative2005}.

\subsubsection{Consumption}

Companies within a broader security industry accrue benefit from perpetuating a state of insecurity~\cite{krahmannMarketOntologicalSecurity2018}. This may be exploited through a narrative of unforeseeable risks that builds uncertainty~\cite{krahmannMarketOntologicalSecurity2018}, fear stimulating consumption~\cite{baudrillardConsumerSocietyMyths1998} in the same way as war~\cite{walkerOutsideInternationalRelations1993}. The security and defence industries need there to be something to be defended against~\cite{neocleousCritiqueSecurity2008}. As well as cyber security being one of these industries, with cyber-security crises leading to increases in budgets~\cite{TalkTalkCISOGiven}, even in other, unaffected, businesses~\cite{havakhorCybersecurityTalentsValue2019}, cyber capability is a factor in being ‘allowed to play’ in others, and can be a barrier to entry. Cyber security is thus enmeshed with a much broader aspect of modernity in the sense of continued consumption, both as an industry in its own right and as a facet of other industries. Some organisations in this study indicated the benefit they accrued from governmental spending on security-related products and services, further demonstrating the link between societal threat and certain business sectors. The Leviathan, which seeks continued and never-ending growth, can stimulate expansion by creating a need for spending that counteracts fear and anxiety~\cite{baudrillardConsumerSocietyMyths1998}. Such spending can arguably “in some way ‘improve’ life in civil society”~\cite[p. 111]{walkerOutsideInternationalRelations1993}. Where that fear and anxiety is generated by unseen and ever-more-sophisticated sources, there is, in theory, no upper limit to the growth that could be achieved.

\section{Conclusion}\label{sec:conclusion}
This paper has shown that Hobbes' work provides a useful lens through which to view the role that cyber security plays in society within and without businesses, particularly given the importance of Hobbesian thinking to Western political thought and the enmeshed nature of states and corporations. Cyber security offers a useful mechanism from which the Leviathan derives benefit. It supports the establishment of fear and discipline, therefore, cementing power through obedience and conformance. Additionally, although less obviously, it also drives accumulation of capital through consumption of products and services, and job creation.

Businesses play a crucial role for the Leviathan. They employ and educate citizens, inuring them to surveillance and punishing them when they transgress. They maintain narratives of morality. They generate and expand capital. In some cases, they operate critical infrastructure and perform other state functions on the Leviathan's behalf. Businesses are themselves mini-Leviathans, and are in fear of threats to their existence. Cyber security functions within those businesses provide a means by which they seek to avoid a state of nature. They also, indirectly, provide that function to the state, supporting its attempts to dominate competing state Leviathans. Actions taken by businesses in relation to cyber security involve spending that provides fuel for the continued growth of the Leviathan's power, and that of the hegemony that the Leviathan supports. 

Our research opens up a number of interesting future directions. The Hobbesian perspective that we introduce may encourage the application of broader contexts from International Relations (IR), using other IR theorists and perspectives to explore cyber security within businesses and wider society. In particular, we consider there to be benefit to research avenues relating to the context of businesses within globalisation and geopolitics, including the establishment of cyber norms and other developing areas of study. Most significantly, our research encourages greater reflexivity within the discipline of cyber security, both for researchers and for practitioners. The latter may benefit from considering the role they unwittingly perform in the maintenance of both political and commercial power structures. Businesses themselves may benefit from reflecting on the role that they play in supporting the state Leviathan, and indeed in wider globalisation of Hobbesian models of control. Such reflexivity and improved awareness of broader contexts offers empowerment for individuals within those businesses, particularly CISOs in this case. Part of the role of a CISO is being the agent of two Leviathans, the mini-Leviathan of the business that employs them, and the larger Leviathan of the state. The latter role may not be as immediately obvious or recognisable and may be uncomfortable for many CISOs. However, recognising this allows for reflexive consideration and engagement with the implications, or at the very least an acknowledgement of what the CISO does or does not agree with. The CISO may acknowledge that there is value in implementing surveillance in order to protect the business-as-Leviathan but if that helps to normalise surveillance by the state Leviathan on the employee-citizen, there may be a potential internal conflict that they need to either resolve or come to terms with. We have considered whether this perspective could potentially lead to (perceived) negative outcomes for the cyber-security industry. For example, will there be less surveillance within organisations as a result of CISOs resisting the support that they indirectly provide to state surveillance? We consider this to be unlikely, however, it opens up further avenues of potential research, particularly to explore whether or not a CISO's primary focus  is on their employer. After all, their employer is offering them the most immediate protection, in terms of salary and continued employment, and, similar to the Hobbesian family structure, is where their primary interest and indeed obedience may lie. Therefore, the CISO may do what is in the interest of their employer and ensure that they protect them as best they can, including implementing controls that they would feel less comfortable with if they were in place in wider society. Their employer's interests may also be more closely aligned with their own, particularly in terms of ensuring continued viability, than with the state Leviathan's, and this symbiotic relationship may be an important factor in any decision-making regarding controls. Future research could also consider the possibility of ‘deviant’ corporations and the concept of insider threats through a Hobbesian lens.\footnote{We thank an anonymous reviewer for these suggestions.} The latter, in particular, could also be expanded to consider the perspective of resistance.\footnote{This has potential parallels with prior work performed by Coles-Kemp et al.~\cite{coles-kempWhyShouldCybersecurity2018}.} Further, there may be an interesting parallel to explore with regard to biological threats, such as Covid-19, and cyber threats. Each of these can be considered as unseen, ephemeral threats that require specialist advisors and motivate restrictive controls that could be considered as offering additional benefits to a Hobbesian state.\footnote{Again, we thank an anonymous reviewer for inspiring this angle.} 

Our Findings offer a broad perspective on cyber-security practice within organisations. As such, they will also be interpreted through other analytical lenses and connected to other, existing, models of managing cyber security.

\bibliographystyle{plain}
\bibliography{local}

\newpage

\appendix

\section{Interview details}\label{sec:interviews}
Below in Table~\ref{tab:industries} we provide a brief summary of the industry sectors represented in this study, whereas Table~\ref{tab:interviews} provides details of the participant interviews.

\begin{table}[h!]
  \footnotesize
  \caption{Industry sectors represented in this study}\label{tab:industries}
  \begin{tabular}{ll @{\hskip 2em} lll}
    \emph{ICB Super-sector} & \emph{Number of organisations} \\
    \midrule    

Banks & 1 \\
Food, Beverage and Tobacco  & 1 \\
Industrial Goods and Services & 6 \\
Personal Care, Drug and Grocery Stores & 2 \\
Real Estate & 1 \\
Technology & 1 \\
Telecommunications & 2 \\
Travel and Leisure & 1 \\
Utilities & 3 \\
    \midrule   
 \textbf{Total} & \textbf{18}
     \end{tabular}
  
  \vspace{1em}
Coverage of industries represented in this research based on classifications taken from~\cite{IndustryClassificationBenchmark2018,CompaniesSecuritiesLondon}.
\end{table}

\vspace{1em}

\begin{table}[h!]
	\footnotesize
	\caption{Participants \& Interviews}\label{tab:interviews}
	\begin{tabular}{ll @{\hskip 2em} lll}
		\multicolumn{2}{c}{\textbf{Participants}} & \multicolumn{3}{c}{\textbf{Interview}}\\
		\emph{ID} &  \emph{Duration} & \emph{Medium} & \emph{Timing} \\
		\midrule    
		CISO1 & 00:48:29 & F2F & Oct19 \\
		CISO2 & 00:49:28 & F2F & Oct19 \\
		CISO3 & 00:47:33 & F2F & Dec19 \\
		CISO4 & 00:44:41 & F2F & Dec19 \\
		CISO5 & 00:43:44 & F2F & Dec19 \\
		CISO6 & 00:41:38 & F2F & Jan20 \\
		CISO7 & 00:45:19 & F2F & Jan20 \\
		CISO8 & 00:49:41 & F2F & Mar20 \\
		CISO9 & 00:51:30 & F2F & Mar20 \\
		CISO10 & 00:38:43 & Remote & Apr20 \\
		CISO11 & 00:55:45 & Remote & May20 \\
		CISO12 & 00:40:56 & Remote & May20 \\
		CISO13 & 00:40:07 & Remote & Jun20 \\
		CISO14 & 00:46:07 & Remote & Jul20 \\
		CISO15 & 00:50:02 & Remote & Jul20 \\
		CEO1 & 00:24:59 & F2F & Dec19 \\
		CEO2 & 00:42:45 & F2F & Jan20 \\
		CFO1 & 00:45:41 & F2F & Jan20 \\
		CFO2 & 00:40:52 & Remote & Apr20 \\
		CIO1 & 00:47:28 & Remote & Jul20 \\
		NED1 & 00:27:52 & F2F & Dec19 \\
	\end{tabular}
	
	\vspace{1em}
	In addition to 15 CISOs, we conducted semi-structured interviews with two Chief Executive Officers (CEOs), two Chief Financial Officers (CIOs), one Non-Executive Director (NED) and one Chief Information Officer (CIO), between October 2019 and July 2020. 
\end{table}

\end{document}